# Metasurface-Inspired Maintenance-Free IoT Tags Characterised in Both Frequency and Time Domains

**Masaya Tashiro[1], Ashif Aminulloh Fathnan[1], Yuta Sugiura[2], Akira Uchiyama[3], and Hiroki Wakatsuchi[1]**

[1] *Department of Engineering, Graduate School of Engineering, Nagoya Institute of Technology, Nagoya 466-8555, Japan*
[2] *Graduate School of Science and Technology, Keio University, Yokohama 223-8522 Japan*
[3] *Graduate School of Information Science and Technology, Osaka University, Suita 565-0871, Japan*

Email: wakatsuchi.hiroki@nitech.ac.jp

We present metasurface-inspired maintenance-free IoT tags that can be characterised not only by frequency-domain profiles but also by time-domain profiles. In particular, time-domain characterisation is made possible by implementing the waveform-selective mechanisms of recently developed circuit-based metasurfaces that behave differently, even at the same frequency, in accordance with the pulse duration of the incident wave. Our designs are numerically and experimentally validated and potentially contribute to accommodating an increasing number of IoT tags within a single wireless network while reducing maintenance effort.

*Introduction:* In recent years, wireless communication technologies have been introduced into a wide variety of devices, instruments and facilities thanks to the development of the Internet of Things (IoT) [1]–[3], which has had an ever-increasing impact on daily life, industries and the economy. For instance, IoT devices enable us to sense objects or collect multidimensional information including variations in space and time [1]–[3]. However, the greater the number of devices implemented to obtain an extensive data set is, the greater the amount of effort needed to maintain the entire system. Usually, IoT devices or sensors send collected information, which consumes energy and thus requires energy storage batteries. One important issue is determining how such a large number of devices can be realistically maintained, including the replacement of batteries for long-term use. Potential solutions may be found in reducing energy dissipation and designing large-capacity batteries; however, these methods are not suitable for low-cost IoT applications such as IoT tags, which are employed to recognise user identification numbers (IDs) or gestures [4]. Hence, it is more ideal to remove the batteries while still maintaining the corresponding sensing mechanism. This may be achieved by observing the frequency profiles of waves scattered from artificially engineered structures called metasurfaces [5]–[8]. Metasurfaces are widely known to be capable of controlling electromagnetic waves at will through use of subwavelength unit cells. In this case, their frequency characteristics can be associated with the IDs of IoT tags. However, the use of frequency-domain characteristics may not be sufficient for accommodating an increasing number of IoT tags. As a potential solution, a series of recent studies reported that circuit-based metasurfaces vary their electromagnetic response, even at the same frequency, depending on the incident waveform or pulse width [9]–[12]. The pulse-width selectivity of such waveform-selective metasurfaces has been exploited to increase the number of degrees of freedom to address issues in antennas [13], [14], electromagnetic compatibility [9], [12], wireless communications [11], [15], signal processing [16], etc. Similarly, the use of the pulse width dimension potentially leads to an increase in the number of maintenance-free IoT tags within a single wireless network. For this reason, this study develops prototypes of metasurface-inspired maintenance-free IoT tags that can be characterised not only by frequency but also by pulse width related to time-domain profiles. We present two types of maintenance-free IoT tag designs that are demonstrated numerically and experimentally. The first design is frequency dependent and is characterised by frequency only. However, the second design has an additional degree of freedom associated with pulse width. Eventually, the second design is experimentally tested in both wired and wireless measurement setups.

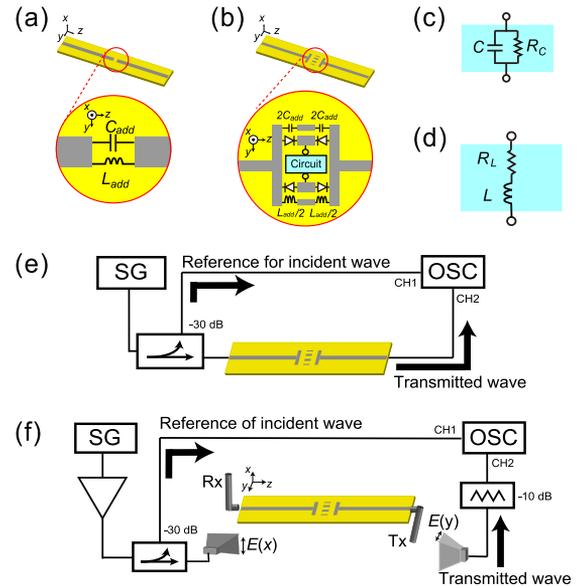

**Fig. 1** *Proposed metasurface-inspired maintenance-free IoT tags. (a) First design and (b) second design containing (c, d) additional circuit components. (c) and (d) represent C-based and L-based circuits, respectively. Measurement setups in (e) wired and (f) wireless environments.*

*Theory and method:* Our IoT tag designs were inspired by recently developed circuit-based metasurfaces or so-called waveform-selective metasurfaces. The waveform-selective mechanisms are explained in detail in the literature [9]–[11], [13], [15], [16]. However, the mechanisms can be briefly described as follows. First, as shown in Figure 1a, our IoT tags are based on an ordinary microstrip design (3.05 mm wide on a 1.52-mm-thick Rogers3003 substrate) that ideally transmits any incoming signal in the frequency of interest, specifically between 2 and 4 GHz. In addition, we implement a pair of a capacitors $C_{add}$ and an inductor $L_{add}$ into the middle of the microstrip, which enables us to set a stop band frequency $f_s$ depending on the values of $C_{add}$ and $L_{add}$ through

$$f_s = 1/2\pi\sqrt{L_{add}C_{add}}. \qquad (1)$$

Moreover, in the second design shown in Figure 1b, the gap bridged by $C_{add}$ and $L_{add}$ is also connected by a diode bridge that contains a paired resistor $R_C$ or $R_L$ and a capacitor $C$ or an inductor $L$, as illustrated in Figures 1c and d. The use of these additional circuit components leads to variable selectivity in the pulse width dimension or in the time domain. Notably, at the resonant frequency $f_s$ determined by $C_{add}$ and $L_{add}$, the diode bridge fully rectifies an incoming signal and converts most of the energy to the zero-frequency component [9], [11], [12]. Therefore, the circuit-embedded microstrip exhibits transient responses influenced by $R_C$, $R_L$, $C$, $L$ and the resistive component of the diodes $R_0$, specifically, [17]

$$\tau_C = CR_CR_d/(R_C + R_d) \qquad (2)$$
$$\tau_L = L/(R_L + R_d) \qquad (3)$$

where $R_d = 2R_0$ and $\tau_C$ and $\tau_L$ denote the time constants of the $C$-based microstrip (containing $R_C$ and $C$) and the $L$-based microstrip (containing $R_L$ and $L$), respectively.

We simulated these microstrip models using the co-simulation method available from ANSYS Electronics Desktop [9], [11], [12], [18]. The measurement sample was fabricated using the design parameters of the simulation models shown in Figures 1a and b. As illustrated in Figure 1e, incident signals were generated by a signal generator (Anritsu, MG3692C) via coaxial cables and a coupler (ET Industries, C-058-30) that sent a portion of the energy to an oscilloscope (Keysight Technologies, DSOX6002A) to later be compared to the transmitted waveform. SMA jacks were used to connect the coaxial cable from the coupler to the microstrips (i.e., the measurement samples). The other sides of the samples were also connected to the oscilloscope via SMA jacks and a coaxial cable. Circuit components were soldered to the measurement samples. We used commercial Schottky diodes provided by Avago (HSMS-286X series). Note that in simulations these diodes were modelled using their SPICE parameters. Moreover, we experimentally tested these samples in the wireless environment shown in Figure 1f. In this case, both ends of the microstrip samples were connected to monopoles (Linx Technologies, ANT-2.4-LCW-SMA) that were oriented to be orthogonal to each other, which reduced the direct propagation between the monopoles as well as between horn antennas (Schwarzbeck Mess-Elektronik, BBHA 9120 D). In this measurement, we also used an amplifier (Ophir, 5193RF) to sufficiently increase the input power (specifically, to 40 dBm) and rectify the incident wave within diode bridges.

*Results:* First, simulation results were obtained for the first microstrip design composed of only $C_{add}$ and $L_{add}$ (Figure 1a), as plotted in Figure 2a. This figure shows that the stop bands from the numerical simulations were almost consistent with the analytical values obtained with Equation (1). Practically, the locations of the stop bands can be associated with the IDs of IoT tags.

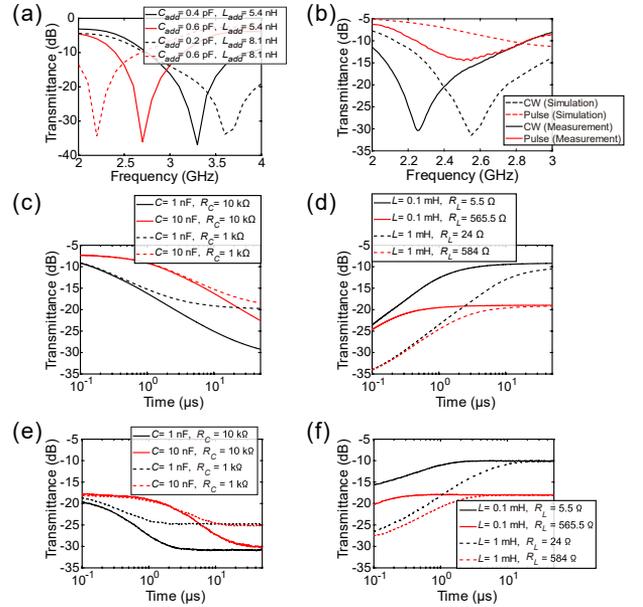

**Fig. 2** *Results with the wired setup (Figure 1e). (a) Simulated frequency-domain profiles of the first design (Figure 1a). (b) Simulated and measured frequency-domain profiles of the second design (Figure 1b) using the C-based microstrip (Figure 1c). Simulated time-domain profiles of (c) the C-based microstrip and (d) the L-based microstrip (Figure 1d) at 2.55 GHz. (e) Measured time-domain profiles of (e) the C-based microstrip and (f) the L-based microstrip at 2.28 GHz.*

Next, we introduced an additional $C$-based circuit with $C_{add}$ and $L_{add}$ fixed at 0.2 pF and 5.4 nH, respectively (Figures 1b and c). This frequency-domain profile was numerically obtained and is plotted in Figure 2b. This figure demonstrates that at the same frequency of 2.55 GHz, the transmittance for a continuous wave (CW) was significantly reduced compared to that for a 50-ns short pulse. This was because the energy of the short pulse was transmitted through the diode bridge, while the CW signal fully charged the internal capacitor $C$, which led to a reduced magnitude of transmittance, as shown in Figure 2a for the first microstrip design.

Due to the above waveform-selective mechanism, the second microstrip design (Figure 1b) provided an additional time-domain profile in Figure 2c. In this figure, the incoming frequency was fixed at 2.55 GHz with a power level of 10 dBm, while $C$ and $R_C$ were varied. In this case, the transmittance was gradually decreased by the waveform-selective mechanism. Notably, Figure 2c indicates that such time-domain characteristics can be properly obtained by adjusting the values of $C$ and $R_C$. For instance, the transition from low to high transmittance was shifted to a larger time scale by increasing $C$. Note that this scale was also consistent with the time constant predicted with Equation (2). In addition, Figure 2c indicates that the transmittance level at steady state was determined by $R_C$, as it relates to the steady-state impedance.

We also simulated the $L$-based microstrip, as shown in Figure 2d. In this case, the transmittance gradually increased in the time domain, as opposed to the trend observed for the $C$-based microstrip, since the electromotive force of $L$ prevented incoming electric charges during the initial time period. However, this force almost disappeared at steady state due to the presence of the rectified zero-frequency component. In addition, the time constants of the $L$-based microstrip displayed good matching with those estimated with Equation (3).

These $C$- and $L$-based microstrips were experimentally validated. The measured frequency-domain profiles were almost consistent with the simulated profiles (Figure 2b). Note that a frequency shift appeared in the solid black curve of Figure 2b in part due to solder that produced additional parasitic parameters and influenced the determination of the resonant frequency in Equation (1). Next, the time-domain transmittances of the $C$- and $L$-based microstrips were measured, as shown in Figures 2e and f, respectively. These figures also indicate that the time-domain characteristics can be changed, as shown in the simulation results in Figures 2c and d.

Finally, we assessed these samples in the wireless environment shown in Figure 1f, and the results are plotted in Figures 3a and b. Overall, the result was the same as that in Figures 2e and f with only minor differences. For instance, the transmittances were entirely reduced since the measurements were conducted in free space.

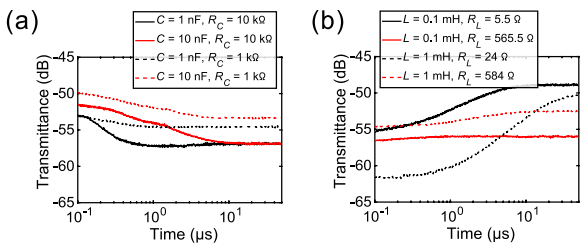

**Fig. 3** *Measurement results for the time-domain profiles of the second design with the wireless setup (Figure 1f): (a) C- and (b) L-based microstrips.*

*Conclusion:* We have presented metasurface-inspired maintenance-free IoT tags that can be characterised and identified not only by frequency-domain profiles but also by time-domain profiles. The proposed designs were based on a typical microstrip with additional circuit components, through which pulse-width-selective characteristics were exploited for use in the time domain. Our study is expected to contribute to accommodating an increasing number of IoT tags within a single wireless network while reducing maintenance effort.

*Acknowledgements:* This work was supported in part by the Japan Science and Technology Agency (JST) under the Precursory Research for Embryonic Science and Technology (PRESTO) Project Nos. JPMJPR193A, JPMJPR1932 and JPMJPR2134.

*Conflict of interest:* The authors declare no conflict of interest.

**References**
[1] Stankovic, J. A.: Research directions for the internet of things. *IEEE Internet Things* **1**(1), 3–9 (2014)
[2] Zanella, A., Bui, N., Castellani, A., Vangelista, L., Zorzi, M.: Internet of things for smart cities. *IEEE Internet Things* **1**(1), 22–32, (2014)
[3] Shafique, K., Khawaja, B. A., Sabir, F., Qazi, S., Mustaqim, M.: Internet of things (IoT) for next-generation smart systems: a review of current challenges, future trends and prospects for emerging 5G-IoT scenarios. *IEEE Access* **8**, 23022–23040 (2020)
[4] Conati, C., et al.: Hand gesture recognition and virtual game control based on 3D accelerometer and EMG sensors. *Proc. 14th Int. Conf. Intelligent User Interfaces* 401–406 (2009)
[5] Yu, N., et al.: Light propagation with phase discontinuities: generalized laws of reflection and refraction. *Science* **334**(6054), 333–337 (2011)
[6] Pfeiffer, C., Grbic, A.: Metamaterial Huygens' surfaces: tailoring wave fronts with reflectionless sheets. *Phys. Rev. Lett.* **110**(19), 197401 (2013)
[7] Yu, N., Capasso, F.: Flat optics with designer metasurfaces. *Nat. Mater.* **13**(2), 139–150 (2014)
[8] Wakatsuchi, H., Greedy, S., Christopoulos, C., Paul, J.: Customised broadband metamaterial absorbers for arbitrary polarisation. *Opt. Express* **18**(21), 22187 (2010)
[9] Wakatsuchi, H., Kim, S., Rushton, J. J., Sievenpiper, D. F.: Waveform-dependent absorbing metasurfaces. *Phys. Rev. Lett.* **111**(24), 245501 (2013)
[10] Eleftheriades, G. V.: Protecting the weak from the strong. *Nature* **505**(7484), 490–491 (2014)
[11] Wakatsuchi, H., Anzai, D., Rushton, J. J., Gao, F., Kim, S., Sievenpiper, D. F.: Waveform selectivity at the same frequency. *Sci. Rep.* **5**(1), 9639 (2015)
[12] Wakatsuchi, H., Long, J., Sievenpiper, D. F.: Waveform selective surfaces. *Adv. Funct. Mater.* **29**(11), 1806386 (2019)
[13] Vellucci, S., Monti, A., Barbuto, M., Toscano, A., Bilotti, F.: Waveform-selective mantle cloaks for intelligent antennas. *IEEE Trans. Antennas Propag.* **68**(3), 1717–1725 (2020)
[14] Barbuto, M., Lione, D., Monti, A., Vellucci, S., Bilotti, F., Toscano, A.: Waveguide components and aperture antennas with frequency- and time-domain selectivity properties. *IEEE Trans. Antennas Propag.* **68**(10), 7196–7201 (2020)
[15] Ushikoshi, D., et al.: Experimental demonstration of waveform‐selective metasurface varying wireless communication characteristics at the same frequency band of 2.4 GHz. *Electron. Lett.* **56**(3), 160–162 (2020)
[16] Imani, M. F., Smith, D. R.: Temporal microwave ghost imaging using a reconfigurable disordered cavity. *Appl. Phys. Lett.* **116**(5), 054102 (2020)
[17] Asano, K., Nakasha, T., Wakatsuchi, H.: Simplified equivalent circuit approach for designing time-domain responses of waveform-selective metasurfaces. *Appl. Phys. Lett.* **116**(17), 171603 (2020)

[18] Homma, H., Akram, M. R., Fathnan, A. A., Lee, J., Christopoulos, C., Wakatsuchi, H.: Anisotropic impedance surfaces activated by incident waveform. *Nanophotonics* **11**(9), 1989-2000 (2022)